\documentstyle[11pt]{article}
\normalbaselines
\begin{document}
\bibliographystyle{unsrt}

\def\bea*{\begin{eqnarray*}}
\def\eea*{\end{eqnarray*}}
\def\ba{\begin{array}}
\def\ea{\end{array}}
\count1=1
\def\be{\ifnum \count1=0 $$ \else \begin{equation}\fi}
\def\ee{\ifnum\count1=0 $$ \else \end{equation}\fi}
\def\ele(#1){\ifnum\count1=0 \eqno({\bf #1}) $$ \else \label{#1}\end{equation}\fi}
\def\req(#1){\ifnum\count1=0 {\bf #1}\else \ref{#1}\fi}
\def\bea(#1){\ifnum \count1=0   $$ \begin{array}{#1}
\else \begin{equation} \begin{array}{#1} \fi}
\def\eea{\ifnum \count1=0 \end{array} $$
\else  \end{array}\end{equation}\fi}
\def\elea(#1){\ifnum \count1=0 \end{array}\label{#1}\eqno({\bf #1}) $$
\else\end{array}\label{#1}\end{equation}\fi}
\def\cit(#1){
\ifnum\count1=0 {\bf #1} \cite{#1} \else 
\cite{#1}\fi}
\def\bibit(#1){\ifnum\count1=0 \bibitem{#1} [#1    ] \else \bibitem{#1}\fi}
\def\ds{\displaystyle}
\def\hb{\hfill\break}
\def\comment#1{\hb {***** {\em #1} *****}\hb }

\newcommand{\TZ}{\hbox{T\hspace{-5pt}T}}
\newcommand{\MZ}{\hbox{I\hspace{-2pt}M}}
\newcommand{\ZZ}{\hbox{Z\hspace{-3pt}Z}}
\newcommand{\NZ}{\hbox{I\hspace{-2pt}N}}
\newcommand{\RZ}{\hbox{I\hspace{-2pt}R}}
\newcommand{\CZ}{\,\hbox{I\hspace{-6pt}C}}
\newcommand{\PZ}{\hbox{I\hspace{-2pt}P}}
\newcommand{\QZ}{\hbox{I\hspace{-6pt}Q}}
\newcommand{\HZ}{\hbox{I\hspace{-2pt}H}}
\newcommand{\EZ}{\hbox{I\hspace{-2pt}E}}
\newcommand{\GZ}{\,\hbox{l\hspace{-5pt}G}}

\vbox{\vspace{38mm}}
\begin{center}
{\bf THE ALGEBRAIC STRUCTURE OF THE 
ONSAGER ALGEBRA}\footnote{Presented at the 8th
Colloquium "Quantum groups and integrable
systems", Prague, 17-19 June 1999.} \\[5mm]

Etsuro Date \footnote{Supported in part
by the Grant-in-Aid for Scientific Research
(B)(2)09440014 Japanese Ministry of Education, 
Science, Sports and Culture}\footnote{E-mail:
ed@math.sci.osaka-u.ac.jp}\\
{\it Department of Mathematics, Osaka
University , Osaka 560, Japan} \\
Shi-shyr Roan
\footnote{Supported in part by the 
NSC grant of Taiwan.}\footnote{E-mail:
maroan@ccvax.sinica.edu.tw}  \\
{\it Institute of Mathematics ,
Academia Sinica ,  Taipei , Taiwan  } 

\end{center}

\begin{abstract}
We study the Lie algebra structure of 
the Onsager algebra from the ideal theoretic 
point of view. A structure
theorem of ideals in the  Onsager algebra is 
obtained  with the  connection to the
finite-dimensional representations.  We  also
discuss the solvable algebra aspect of the
Onsager algebra through the formal
Lie algebra theory.
\end{abstract}

\section{Introduction}
In this note, we are going to discuss the 
algebraic structure of a certain
infinite-dimensional Lie algebra which appeared in
the seminal work of Onsager of 1944 on the
solution of  2D Ising model \cite{O}. 
Due to  some other simpler and 
powerful methods introduced later in the study of
Ising model, the algebra in the Onsager's original
work, now called the  
 Onsager algebra, had not 
received enough attention in the substantial
years until the 1980-s 
when  its new-found role appeared in the 
superintegrable chiral Potts model,
[2-12]  
(for a review of the subject, see \cite{UI} and
references therein ). 
Since the work of Onsager it was
known that there exists an intimate relationship
between the Onsager algebra and $sl_2$, 
a connection  
now clarified in \cite{D, R91}.
Namely, the Onsager algebra
is isomorphic to the fixed subalgebra of 
$sl_2$-loop algebra (or alternatively, its
central  extension $A_1^{(1)}$, ) by a certain
involution.  A
generalization of the Onsager algebra to other 
Kac-Moody algebras was later introduced in 
\cite{UI, AS}, with  some interesting relations
with integrable motions.  
Due to the close relationship of the
Onsager algebra  with 2D integrable 
models in statistical mechanics, especially in the
chiral  Potts model, a thorough 
mathematical understanding of 
 this infinite-dimensional algebra is  
desirable to warrant the further
investigation. In this note, we summarize our
recent results on the algebraic study of the 
structure of  the Onsager algebra.  Detailed
derivations, as well as extended references to
the literature, may be found in 
\cite{DR}. We have established the
structure theorem of a  certain class of
(Lie)-ideals of the Onsager  algebra,
originated from the study of (reducible 
or irreducible) finite dimensional
representations of the Onsager algebra, 
through the theory of the 
reciprocal polynomials. 
In the process, we have found a 
 profound structure of the Onsager algebra 
 with some interesting applications
to solvable and nilpotent
Lie algebras. It suggests that the further study
of the ideals in the Onsager algebra and
its possible generalization  should provide
certain useful information on the theory of
solvable Lie algebras.  The mathematical results
obtained would be expected to  have some feedback
to  the original physical theory. Such a program
is now under progress and partial results are
promising.

\section{Hamiltonian of Superintegrable Chiral Potts Model}
The  superintegrable chiral Potts
$N$-state  spin chain Hamiltonian  has the 
following form of a  
parameter $k'$,
$$
H(k') = H_0 + k' H_1 \ , 
$$
with $H_0, H_1$ the 
Hermitian operators
acting on the vector space of 
$L$-tensor of
$\CZ^N$, defined by  
$$
H_0 = -2 \sum_{l=1}^L \sum_{n=1}^{N-1} (1-\omega^{-n})^{-1}
X_l^n , \ \  \  
 H_1 = -2 \sum_{l=1}^L \sum_{n=1}^{N-1} (1-\omega^{-n})^{-1}
Z_l^nZ_{l+1}^{N-n} \ ,
$$
where $\omega = e^{\frac{2 \pi
{\rm i}}{N}} $,  
$
X_l = I \otimes \ldots \otimes \stackrel{l{\rm
th}}{ X}
\otimes \ldots \otimes I $, $ 
Z_l = I \otimes \ldots \otimes \stackrel{l {\rm
th}}{ Z}
\otimes \ldots \otimes I ,  (Z_{L+1}= Z_1 ) $.  
Here $I$ is the identity operator, and  
$X, Z$ are the operators of $\CZ^N$ with 
the Weyl
commutation  relation, $ZX = \omega XZ$, which are
defined by $X|m>= |m+1> , Z|m> = \omega^m |m>$,  
$m \in \ZZ_N$. The operator
$H(k')$ is  Hermitian for real $k'$, hence
with the real eigenvalues. 
When $N=2$,  $X, Z$, become the
Pauli matrices,  
$\sigma^1, \sigma^3$, then it is 
the Ising quantum chain \cite{K}:
$$
\begin{array}{l}
-H(k') = \sum_{l=1}^L \sigma^1_l 
+ k' \sum_{l=1}^L \sigma_j^3 \sigma_{l+1}^3 \ .
\end{array}
$$
For $N=3$, one obtains the 
$\ZZ_3$-symmetrical self-dual chiral clock model 
with the chiral angles $\varphi= \phi =
\frac{\pi }{2}$,    
$$
\frac{- \sqrt{3}}{2}H(k') = \sum_{l=1}^L
(e^{\frac{-
\pi {\rm i}}{6}} X_l + e^{\frac{\pi
{\rm i}}{6}} X_l^2 ) + k' \sum_{l=1}^L 
(e^{\frac{- \pi
{\rm i}}{6}} Z_lZ^2_{l+1} + e^{\frac{\pi
{\rm i}}{6}} Z_l^2Z_{l+1}) \ ,
$$
which was studied by  
Howes, Kadanoff and M. den
Nijs \cite{HKN}. Note that 
the Potts model is given by the chiral
angles, $\varphi=
\phi = 0$. 
For a general $N$,
$H(k')$ was constructed in a paper of 
G. von Gehlen and R. Rittenberg \cite{GR},
in which the models were shown to 
be  "superintegrable" in the sense that the
Dolan-Grady 
(DG) condition \cite{DG} is satisfied for 
$A_0 = -2 N^{-1} H_0 , A_1
= -2 N^{-1} H_0 $, 
$$
[A_1, [A_1, [A_1, A_0]]]= 16 [A_1, A_0] \ , \ \ 
[A_0, [A_0, [A_0, A_1]]]= 16 [A_0, A_1] \ .
$$
For a pair of operators, $A_0, A_1$, we denote, $
4G_1 =  [A_1, A_0 ]$, and  
define an infinite sequence of operators, 
$A_m , G_m, ( m \in \ZZ )$, by 
relations,
$$
A_{m-1} - A_{m+1} = \frac{1}{2} [A_m , G_1 ] \ , \ \ 
G_m = \frac{1}{4} [A_m, A_0 ]  \ .
$$
The DG-condition on  $A_1, A_0,$ are equivalent 
\cite{D, R91} to  the statement that the  
collection of 
$A_m,
G_m$ forms the Onsager algebra, which means they
satisfy the  Onsager relations,
\bea(lll)
 [A_m , A_l ]  = 4 G_{m-l} \ , &
 [A_m , G_l ] = 2 (A_{m-l} - A_{m+l} ) \ ,
& [G_m, G_l] = 0 \ .
\elea(OA)
The above relations ensure that 
the
$k'$-dependence eigenvalues of $H(k')$ have 
the following special form as in the Ising model,
$$
 a  + b k'   + 2N
\sum_{j=1}^{n} m_j 
\sqrt{1+k'^2 - 2k' \cos (\theta_j) }
\ ,
$$ 
where $a, b, \theta_j, $ are reals, and $m_j$'s 
take the values, 
$m_j =  -s_j, (-s_j +1) , \ldots ,  s_j$ 
with $s_j$ a positive half-integer 
\cite{AMP, D}. Mathematically, this
result follows from the  classification of the
finite-dimensional unitary representations of the
Onsager algebra, by the relationship of the
algebra with the loop algebra of
$sl_2$, $L(sl_2):= \CZ[t, t^{-1}] \otimes sl_2$,
which was identified in \cite{R91} as follows.   
\par \vspace{.2in} \noindent
{\bf Definition.} The Onsager algebra, denoted
by $ {\sf OA}$, is defined as one of the following
equivalent conditions:

(i) ${\sf OA}$ = the universal Lie algebra
generated by  two elements,
$A_0, A_1$, with the DG-condition. 

(ii) ${\sf OA}$ = the Lie-subalgebra of
$L(sl_2)$ fixed by  the involution
$\hat{\theta}$, 
$$
\hat{\theta} : \ p(t) \longmapsto p(t^{-1}) \ , \
e
\longmapsto  f, \ f \longmapsto e,  \ h
\longmapsto  -h \  \ , 
$$
where $p(t) \in \CZ[t, t^{-1}]$, $e, f, h$ are
the standard basis of 
$sl_2$ with $
[e, f] = h , \ [h, e] = 2e , \ 
[h, f] = -2f $.  
$\Box$ \par \vspace{.2in} \noindent 
The sequence, $A_m, G_m$, associated to the
universal elements, 
$A_0, A_1$ of (i) have the following expression 
as elements in (ii), 
$$
A_m = 2 ( t^m e + t^{-m} f ) \ , \ \ 
G_m = (t^m - t^{-m}) h \ , \ \ {\rm for} \ \ m \in
\ZZ \ .
$$
We shall always make the above identification in
what follows.  
For an element  $X$ in $L(sl_2)$, the 
criterion of $X$ in ${\sf OA}$ is now given by
$$ 
X \in {\sf OA} \   \Longleftrightarrow  \ \ X = 
p(t)e + p(t^{-1})f +
q(t)h \ , \ \ {\rm with} \  \ q(t)+
q(t^{-1}) = 0 \ ,
$$
where $ p(t) , q(t) \in \CZ[t, t^{-1}]$. 
In fact, $q(t)$ can always be written
in  the form, $
q(t) = q_+(t) - q_+ (t^{-1})$ with 
$q_+(t) \in \CZ[t]$.

\section{Closed Ideals of the Onsager Algebra} 
A (non-trivial Lie) ideal ${\sf I}$ of a Lie
algebra 
${\sf L}$  (over $\CZ$) is called a closed 
ideal if it satisfies the following condition,
$$
{\sf I} = \{ x \in {\sf L} \ | \ [x, {\sf L}]
\subset  {\sf I} \} , 
$$
or equivalently, ${\sf L}/{\sf I}$ has the
trivial  center. By Schur's lemma,  the kernel
ideal of an irreducible representation of ${\sf
L}$ in
$sl_n(\CZ)$ is always closed, which constitutes
an  important class of closed ideals. 
In this section, we shall describe the
classification of closed ideals of the Onsager
algebra, {\sf OA}. 
\par \vspace{.2in} \noindent
{\bf Definition.} 
(i) Let 
$P(t)$ be a non-trivial monic
polynomial in $\CZ[t]$. We call $P(t)$ a 
 reciprocal polynomial if 
$P(t) = \pm t^d P(t^{-1})$, where $d$ is the
degree of $P(x)$.

(ii) For a reciprocal polynomial $P(t)$, 
${\sf I}_{P(t)}$ is the ideal of ${\sf OA}$
defined by
$$
{\sf I}_{P(t)} := \{ X = 
p(t)e + p(t^{-1})f +
q(t)h \in {\sf OA} \ | \
p(t) , q(t) \in P(t)\CZ[t,
t^{-1}]  \} \ .
$$
We shall call $P(t)$ the generating polynomial of
the ideal ${\sf I}_{P(t)}$.
$\Box$ \par \vspace{.2in} \noindent 
It is easy to see that 
zeros of a reciprocal polynomial $P(t)$
not equal to
$\pm 1$ must occur in the reciprocal  pairs, and 
${\sf I}_{P(t)}$ is invariant under the
involution of ${\sf OA}$ , $A_m
\mapsto A_{-m}, G_m \mapsto G_{-m}$. Through the 
Chinese remainder theorem, one can establish the 
a canonical (Lie-)isomorphism of the quotient
algebras,
$$
{\sf OA}/{\sf I}_{P(t)}
\stackrel{\sim}{\longrightarrow}
\prod_{j=1}^J  {\sf OA}/{\sf I}_{P_j(t)} \ , \ \
\ \ P(t) := \prod_{j=1}^J P_j(t)
$$
where $P_j(t)$  are pairwise relatively prime 
reciprocal polynomials. The role of 
reciprocal polynomials in the study of the Onsager
algebra  is given by the following structure 
theorem of closed ideals of ${\sf OA}$.
\par \vspace{.2in} \noindent
{\bf Theorem 1.} An ideal ${\sf I}$ of ${\sf OA}$ 
 is closed if  and only
if ${\sf I} = {\sf I}_{P(t)}$ for a reciprocal
polynomial
$P(t)$ whose zero  at
$t=\pm1$ are of the even multiplicity.
The ideal ${\sf I}_{P(t)}$ 
is characterized as the minimal closed
ideal of ${\sf OA}$ containing 
$P(t)e+ P(t^{-1})f$.   
$\Box$ \par \vspace{.2in} \noindent  
For an irreducible special representation of 
${\sf OA}$ on a finite dimensional vector space 
$V$, $\rho: {\sf OA}
\longrightarrow sl(V)$, the 
generating polynomial of the kernel 
 ${\rm Ker}(\rho)$  is given by 
$P(t) = \prod_{j=1}^n U_{a_j}(t)$, 
where $a_j \in \CZ^*-\{ \pm 1 \}, a_j \neq
a_i^{\pm 1}$ for $j \neq i$, and 
$U_a(t) := (t-a)(t-a^{-1})$. In this situation, 
the evaluation morphism of ${\sf OA}$, 
induced from $L(sl_2)$,  
into the sum of $n$
copies of $sl_2$, 
$$
e_{a_1, \ldots, a_n} : {\sf OA} \longrightarrow
\bigoplus^n sl_2
\ , \ \ 
\ \ X \mapsto ( e_{a_1}(X), \ldots, e_{a_n}(X)) \
,
$$
gives rise the isomorphism between 
${\sf OA}/{\rm Ker}(\rho) $ and $
\stackrel{n}{\bigoplus}sl_2$. Hence 
irreducible representations of the  $sl_2$-factors
determine an irreducible representation of 
${\sf OA}$.

\section{Completion of the Onsager Algebra and
 Solvable  Lie Algebras}
By the previous discussion, the study of
closed ideals in 
${\sf OA}$ can be reduced to the case, 
 ${\sf I} 
= {\sf I}_{P(t)}$ with $P(t)= (t \pm 1)^L$ or $  
U_a(t)^L$ for $L \in \ZZ_{\geq 0} , \ a
\in
\CZ^*-\{\pm 1\}$. In this report, we shall only
discuss the case, $P(t) = (t \pm 1)^L$. As the
map,
$t \mapsto -t$, gives rise an involution of 
${\sf OA}$, one has the isomorphism,
$$
{\sf OA}/{\sf I}_{(t-1)^L} \ \ \simeq \ \ {\sf
OA}/{\sf I}_{(t+1)^L} \ .
$$ 
We may assume,  $P(t) =
(t-1)^L , ( L \in \ZZ_{\geq 0} )$. 
Denote $\pi_L, \pi_{KL}$ the canonical
projections,
$$
\begin{array}{ll}
\pi_L : {\sf OA} \longrightarrow {\sf OA}/{\sf
I}_{(t-1)^L} \ , & 
\pi_{KL} : {\sf OA}/{\sf I}_{(t-1)^L}
\longrightarrow  {\sf OA}/{\sf I}_{(t-1)^K} \ , \ 
L \geq K \geq 0 \ ,
\end{array}
$$
and $\widehat{\sf OA}$ the projective limit of 
the projective system, $({\sf OA}/{\sf
I}_{(t-1)^L}, 
\pi_{KL})$. Then 
$\widehat{\sf OA}$ is a Lie algebra and denote the
canonical morphism, $
\psi_L : \widehat{\sf OA} \longrightarrow 
{\sf OA}/{\sf I}_{(t-1)^L}, L \in \ZZ_{\geq 0}$,
with the kernel, $\widehat{\sf OA}^L : =
{\rm Ker}(\psi_L)$. We have,
$\widehat{\sf OA}/\widehat{\sf OA}^L \ \simeq \ {\sf OA}/
{\sf I}_{(t-1)^L}$. We have a filtration of
ideals in 
$\widehat{\sf OA}$,
$$
\widehat{\sf OA} = \widehat{\sf OA}^0 \supset 
\widehat{\sf OA}^1 \supset \cdots \supset \widehat{\sf
OA}^L \supset \cdots \ \ \ \  . 
$$
There exists a morphism, $
\pi: {\sf OA} \longrightarrow \widehat{\sf OA}$, 
with $\psi_L \pi = \pi_L$. The Lie 
algebra $\widehat{\sf OA}$ is regarded as
a completion of ${\sf OA}$. For the convenience,
we shall write the element
$\pi (X)$ of  $\widehat{\sf OA}$ again by $X$ for 
$X \in {\sf OA}$
if no confusion could arise. 
 There exists the unique sequence of  
 elements, $X_k, Y_k, ( k \in
\ZZ_{\geq 0})$, in $\widehat{\sf OA}$ such that
the following identities hold in $\widehat{\sf
OA}$,
$$
A_m (= \pi (A_m)) =  \sum_{k\geq 0}
\frac{m^{(k)}}{k!} X_k
\ , \ \ 
\ \ G_m (= \pi (G_m)) = \sum_{k\geq 0} 
(-1)^k \frac{m^{(k)}}{k!} Y_k \ , 
$$
where  $x^{(n)}
$ is the shifted factorial defined by
$x^{(0)} := 1, x^{(n)} := x(x-1)
\cdots (x-n+1),  
n \in \ZZ_{> 0}$. 
In fact, with the infinite formal sum 
$\widehat{\sf OA}$ has 
$X_k, Y_k$ as the generators of the  formal
Lie algebra, while $\widehat{\sf
OA}^L$ is  the ideal generated by 
$X_k, Y_k, (k \geq L)$.
In ${\sf OA}/{\sf I}_{(t-1)^L}$, one has,  
$$
\psi_L(A_m) = \sum_{k=0}^{L-1}
\frac{m^{(k)}}{k!} \psi_L (X_k) \ , \ \ \ 
\psi_L(G_m ) = \sum_{k=0}^{L-1}(-1)^k
\frac{m^{(k)}}{k!} \psi_L (Y_k) \ , \ \ m \in
\ZZ \ . 
$$
The relations (\req(OA)) for ${\sf OA}$ now
become the following  relations in $\widehat{\sf
OA}$, 
\begin{eqnarray*}
&\sum_{n, k \geq 0}
\frac{a!b!}{n!k!}
s^n_as^k_b[X_n, X_k ] = 4  
\sum_{k \geq 0} (-1)^{b+k}
\frac{(a+b)!}{k!} s^k_{a+b} Y_k \ , 
\\
&\sum_{n,k} (-1)^n \frac{a!b!}{n!k!} s^n_as^k_b 
 [Y_n , X_k ] = 2
(1-(-1)^a )  \sum_{k}\frac{(a+b)!}{k!} s^k_{a+b}
X_k \ , \\ 
& [Y_n, Y_k ] = 0 \ ,
\end{eqnarray*}
where $s^n_k ,  S^n_k, (n, k \in
\ZZ_{\geq 0}), $ are the Stirling numbers,
i.e., the
integers with the relations, $
x^{(n)}= \sum_{k \geq 0} x^k s^n_k \ ,  \ x^n=
\sum_{k \geq 0} x^{(k)}S^n_k $.
Note that $s^n_n = S^n_n = 1$, and $ 
s^n_k = S^n_k = 0$ for $ k>n $. By the 
identities among Stirling's numbers 
\cite{Ri},
$$
\sum_{k}  s^a_kS^k_b = \delta^a_b  \ ,
\ \ \ \frac{b!}{a!} \sum_{k} (-1)^k s^a_kS^k_b =
(-1)^a  ( \begin{array}{c} a-1 \\ b-1 \end{array})
\ , \ \ 
\frac{a!}{j!} S^j_a= 
\sum_{l} 
\frac{k!(a+l)!}{l! (j+k)!} s^l_kS^{j+k}_{a+l} \ ,
$$
one obtains the commutation relations of 
$X_k, Y_k$,
$$
\begin{array}{ll}
[X_n, X_k ] &= 4 (-1)^n (Y_{n+k} + \sum_{a > 0} 
( \begin{array}{c} a+k-1 \\ a
\end{array} ) Y_{n+k+a} ) \ ,  \\

[Y_n, X_k ] &= 2 (((-1)^n-1) X_{n+k} - \sum_{a >0}
(-1)^a ( \begin{array}{c} a+n-1 \\ a
\end{array} )  X_{n+k+a} ) \ ,  \\

[Y_n, Y_k ] & = 0 \ .
\end{array}
$$
By the above relations,  $\widehat{\sf
OA}^L/\widehat{\sf OA}^{L+1}$ is abelian for a
positive integer $L$, hence 
$\widehat{\sf
OA}/\widehat{\sf OA}^L $ is a 
finite-dimensional solvable Lie-algebra.
However, there exist certain non-trivial
relations  among $Y_k$. In fact, one can  express 
$Y_{2n}$ in terms of $Y_{2k+1}$ for $k
\geq n$ in the algebra $\widehat{\sf OA}$; 
the explicit formula is given by
$$
Y_{2n} = \sum_{k \geq n} (-1)^{k-n+1}
\frac{(4^{k-n+1}-1) B_{k-n+1}}{k-n+1} (\begin{array}{c}
2k \\ 2n-1
\end{array} ) Y_{2k+1} \ ,
$$
where $B_j  (j \geq 1)$ are the Bernoulli
numbers, $B_j = (-1)^{j-1}b_{2j}$, with
$\frac{x}{e^x-1}= 
\sum_{j=0}^{\infty}\frac{b_j}{j!}
x^j$. For the study of the Lie algebra structure 
of $\widehat{\sf OA}$, it is more natural to 
use a local coordinate system near $t=1$ for  
expressing elements in  
$L(sl_2)$. A convenient variable is,  
$\lambda := \frac{t-t^{-1}}{2}$, near the
origin, $\lambda=0$. The structure of
$\widehat{\sf OA}$  can be visualized as a formal
subalgebra of  
$sl_2[[\lambda]] ( = \CZ[[\lambda]]  \otimes 
 sl_2 
)$ as follows.
\par \vspace{.2in} \noindent
{\bf Theorem 2 .} Define 
$$
\begin{array}{ll}
 sl_2 \langle\langle \lambda \rangle\rangle   &:
= 
\CZ[[\lambda^2 ]] h +  
\lambda \CZ[[\lambda^2 ]] e + \lambda \CZ[[\lambda^2 ]]
f \ \subset sl_2[[\lambda]] , \\
sl_2\langle\langle \lambda \rangle\rangle^L &: = 
sl_2\langle\langle \lambda \rangle\rangle \bigcap
\lambda^L sl_2[[\lambda]] \ , \ \ (L \in
\ZZ_{\geq 0}) \ .
\end{array} 
$$
There is a formal 
Lie-algebra isomorphism, 
$$
\widehat{\sf OA} \simeq  sl_2 \langle\langle 
\lambda
\rangle\rangle   \ \ ,  
$$
under which  $\widehat{\sf
OA}^L$ is corresponding to 
$sl_2\langle\langle  \lambda \rangle\rangle^L$.
As a consequence,  
$$
{\sf OA}/{\sf I}_{(t-1)^L} \simeq  sl_2
\langle\langle 
\lambda \rangle\rangle/sl_2
\langle\langle  \lambda \rangle\rangle^L \ .
$$
$\Box$ \par \vspace{.1in} \noindent
By the above Theorem,  ${\sf OA}/{\sf I}_{(t-1)^L}
$ is a solvable Lie algebra of dimension $L+
[\frac{L}{2}]$, which has a basis consisting of 
$\psi_L(X_k),
\psi_L(Y_l), 0
\leq k , l <L, l \equiv 1 \pmod{2}$. By the 
structure of ${\sf OA}/{\sf I}_{(t-1)^L} $, 
one can easily see that 
the criterion of trivial center for ${\sf
OA}/{\sf I}_{(t-1)^L}$ , (or 
equivalently, the closed ideal for ${\sf
I}_{(t-1)^L}$), is given by 
the integer
$L$ to be even. For odd 
$L$, the center of ${\sf OA}/{\sf
I}_{(t-1)^L}$ 
is 1-dimensional.   
Using the relations,
$$
t-1  = \lambda -1 + \sqrt{1+
\lambda^2} \ , \ \
t^{-1}-1  =  - \lambda -1 +
\sqrt{1+
\lambda^2} \ ,
$$
one obtains an explicit expression of $A_m, G_m$, 
in terms of $\lambda$ in the above theorem, in 
particular,
$$
A_0  = 2  (e +   f) \ , \ \ 
A_1  =  
2  (e-f) \lambda + 2  (e +  f) 
\sum_{j \geq 0} ( \begin{array}{c} \frac{1}{2} \\
j\end{array} )  \lambda^{2j} \ .
$$
By taking the expression of 
of $A_0,
A_1$ modular $\lambda^L
sl_2[[\lambda]]$, one can 
construct reducible representations of ${\sf
OA}$  with the kernel ideal generated by
$(t-1)^L$.

\section{ Further Remarks}
For the understanding of 
representations of the Onsager algebra, both
the  reducible and irreducible ones, the
structure of 
${\sf OA}/{\sf I}_{P(t)}$ with a reciprocal
polynomial $P(t)$ warrants the mathematical
investigation. 
But, just to keep things simple, we restrict our
attention in this
present report only to the case, $P(t) = (t \pm
1)^L, L \geq 1$, or $ U_a(t), a \neq \pm 1$. With 
a similar argument, one can also obtain the
structure of quotients for closed ideals
generated by
$P(t) = U_a(t)^L, L \in
\ZZ_{>0}$,  ( for the details, see \cite{DR}). 
The results, together with the mixed types, 
should have some interesting applications and
implications in   solvable or nilpotent algebras, 
which we shall leave to future work.
Here is the one  example which appeared in
\cite{R91}:
\par \vspace{.2in} \noindent
{\bf Example.} ${\sf OA}/{\sf I}_{(t^2+1)^2}
\simeq sl_2[\varepsilon ]/ \varepsilon^2
sl_2[\varepsilon ]$. Denote
$e_k, f_k, h_k, (k=0,1)$ the basis elements of
$sl_2[\varepsilon ]/ \varepsilon^2
sl_2[\varepsilon ]$ corresponding to the standard
basis of 
$sl_2$ with the index $k$ indicating the grade of 
$\varepsilon$. Define the linear map, $
\varphi : {\sf OA} \longrightarrow 
sl_2[\varepsilon ]/ \varepsilon^2sl_2[\varepsilon
]$, by the expression:
$$
\begin{array}{l}
\varphi (A_m) = 2 ( {\rm i}^m e_0 + {\rm i}^{-m} f_0 + 
m {\rm i}^{m-1} e_1 + m{\rm i}^{-m+1}f_1 ) \ , \\ 
\varphi (G_m) = ({\rm i}^m- {\rm i}^{-m})h_0 + 
m({\rm i}^{m-1}-{\rm i}^{m-1})h_1 \ .
\end{array}
$$
Then $\varphi$ is a surjective 
morphism. Since $sl_2[\varepsilon ]/ 
\varepsilon^2sl_2[\varepsilon ]$
has only trivial center, ${\rm Ker}(\varphi)$ is  a
closed ideal with $(t^2+1)^2$ as the generating
polynomial. 
$\Box$ \par \vspace{.2in} \noindent
Generalizations of the Onsager algebra to
other loop  algebras or Kac-Moody algebras as in 
\cite{UI, AS} should provide ample
examples of solvable algebras of  the above type. 
We hope that the further development of the study of 
the subject 
will eventually lead to some 
interesting results in Lie-theory with possible 
applications in quantum integrable models.


\begin{thebibliography}{99}

\bibitem{O} L. Onsager: Phys. Rev. 65 (1944)
117.


\bibitem{AMPT} G. Albertini, B. M. McCoy, J. H.
H. Perk, and S. Tang: Nucl. Phys. B 314 (1989)
741.

\bibitem{AMP} G. Albertini, B. M. McCoy, and 
J. H. H. Perk:  Phys. Lett. A
135 (1989) 159;  Phys.
Lett. A 139 (1989) 204; Adv. Stud. Pure
Math., vol. 19, Kinokuniya Academic 1989 .

\bibitem{AMPTY} H. Au-Yang, B. M. McCoy,
J. H. H. Perk, S. Tang, and M. L. Yan:  Phys.
Lett.  123 A (1987) 219.


\bibitem{B} R. J. Baxter: Phys. Letts. A 133
(1988) 185.

\bibitem{BPA} R. J. Baxter, J. H. H. Perk and H.
Au-Yang:  Phys.  Letts. A 128 (1988)
138.

\bibitem{BBP} R. J. Baxter, V.V. Bazhanov and J.
H. H. Perk: Int. J. Mod.
Phys. B 4 (1990) 803.


\bibitem{D} B. Davies:  J. Phys. A: Math. Gen. 23
(1990) 2245;   J. Math.  Phys.
32 (1991) 2945.

\bibitem{DG} L. Dolan and M. Grady:  Phys. Rev. D
25 (1982) 1587. 

\bibitem{GR} G. von Gehlen and R. Rittenberg,
Nucl. Phys. B 257 (1985) 351.

\bibitem{HKN} S. Howes, L.P. Kadanoff and M. den
Nijs:  Nucl.
Phys. B 215 (1983) 169.


\bibitem{P} J. H. H. Perk: in $ Proc. 
Symp. in Pure Mathematics,$  American
Mathematical Society, 1989, Vol. 49,  part 1 
p. 341. 

\bibitem{UI} D. B. Uglov and I. T. Ivanov:  J.
Stat. Phys. 82  (1996) 87. 

\bibitem{R91} S. S. Roan:{\it Onsager's algebra,
loop algebra and chiral Potts model}, MPI 91-70,
Max-Planck-Institut f\"{u}r Mathematik, Bonn,
1991. 

\bibitem{AS} C. Ahn and K. Shigemoto: Modern
Phys. Lett. A 6 (1991) 3509.

\bibitem{DR} E. Date and S. S. Roan: {\it The
structure of quotients of the Onsager algebra by
closed ideals}, Preprint math.QA/9911018.

\bibitem{K} J. B. Kogus:  Rev. Mod.
Phys. 51 (1979) 659 and references therein.



\bibitem{Ri} J. Riordan: {\it An introduction to
combinatorial analysis}, Wiley publication in
statistics 1958 ; {\it Combinatorial identities},
Wiley series in probabilty and mathematical
statistics 1968. 

\end{thebibliography}
\end{document}